\documentclass[a4paper,10pt]{article}

\usepackage{amssymb,amsfonts,amsmath,epsfig,float,graphicx}
\usepackage {url}

 \newtheorem{theorem}{Theorem}[section]
 \newtheorem{lemma}[theorem]{Lemma}
 \newtheorem{proposition}[theorem]{Proposition}

\def\upd{{\rm d}}
\def\prob{{\rm Prob\,}}
\newcommand{\be}{\begin{eqnarray}}
\newcommand{\ee}{\end{eqnarray}}

\def\Eo{{\mathbb E}}
\def\Io{{\mathbb I}}

\newcommand{\chapter}[1]{\section*{Chapter #1}}

\def\beginproof{\par\strut\vskip 0.5cm\noindent{\bf Proof}\par}
\def\endproof{\par\strut\hfill$\square$\par\vskip 0.5cm}

\begin{document}

\title{Large deviation estimates\\
involving deformed exponential functions}
\author{Jan Naudts and Hiroki Suyari\footnote{On leave of absence from Chiba University.}\\
Universiteit Antwerpen}
\maketitle

\begin{abstract}
We study large deviation properties of probability distributions
with either a compact support or a fat tail by comparing
them with q-deformed exponential distributions.
Our main result is a large deviation property for probability distributions
with a fat tail.
\end{abstract}

\section{Introduction}

The Law of Large Numbers (LLN) states that the arithmetic mean of i.i.d.~variables
$X_1,X_2,\cdots,X_n$ converges to the first moment $\Eo X_k$ of the probability
distribution. The Large Deviation Principle (LDP) is the property that 
the probability that the arithmetic mean has a deviating value is
exponentially small in the number of variables $n$. It is an important assumption
for the theorem of Varadhan \cite {VS66}, which deals with the asymptotic evaluation
of certain integrals. 
See also \cite {ERS85,DS89,DZ98,dHF00,FK06,TH09}.

Varadhan's theorem is a generalization of Laplace's method of evaluating integrals.
As such it is highly relevant for the axiomatic formulation of statistical mechanics. 
The standard reference in this direction is the book of Ellis \cite {ERS85}.
A more recent review is found in \cite {TH09}. 
The breakdown of Varadhan's theorem is related with the occurrence of phase transitions
in models of statistical physics. It is due to the appearance of strong correlations
between the variables $X_k$. Another reason of failure of
Varadhan's theorem can be that the LDP is not satisfied.
This is the case for instance when the probability distribution of the variables $X_k$
has a fat tail. It is the latter situation which is considered in the present work.

Mathematicians have studied large deviations in the context of probability distributions
with a fat tail starting with the works of Heyde \cite{HCC67, HCC67b}
and Nagaev \cite{NAV69,NAV69a}. See also \cite{NSV79,R93,V94,KM97,MN98,STJ01,NTYY04,LL09}.
The present work starts from the question whether a systematic use of
so-called q-deformed exponential functions can make a contribution to this area of research.
The q-deformed exponential functions, used in the present work, have been introduced \cite{TC94}
in the context of non-extensive statistical physics \cite{TC88}. See also \cite{TC10,NJ11}.
Our approach differs from that of \cite {RLT12} and of \cite{SS14}
who consider strong correlations in the 
context of nonextensive statistical mechanics.

The strategy of the paper is to mimic the standard approach, replacing 
where meaningful the exponential function by a deformed function.
We therefore start in the next section by reviewing some standard inequalities.
Section 3 gives the definition of q-deformed exponential and logarithmic functions.
Section 4 deals with an application of the Markov inequality in the case
of distributions with a compact support.
The treatment of distributions with a fat tail is more difficult.
Before discussing them in Section 6 we first study the q-exponential distributions
in Section 5. The final Section 7 contains a summary and an evaluation of what
has been obtained.

\section{The standard inequality}

The Markov inequality
\be
\prob\left(X\ge x\right)\le \frac {\Eo X}x,
\quad x>0,
\ee
valid for any random variable $X$ assuming non-negative values,
implies that for any random variable $X$ which assumes real values
one has
\be
\prob\left(X\ge x\right)\le A(a)e^{-a x},
\quad a\ge 0.
\label{intro:ineq}
\ee
This expression involves the moment generating function
\be
A(a)=\Eo e^{a X}.
\ee
Its existence is called Cram\'er's condition.
For a sequence $X_1,X_2,\cdots,X_n$ of i.i.d.~variables
there follows
\be
\prob\left(\frac 1n\sum_{k=1}^nX_k\ge x\right)\le A^n(a)e^{-na x}.
\label{intro:ub}
\ee
Introduce a {\sl rate function} $I(x)$ defined
by
\be
I(x)=\sup_{\theta\ge 0}\{\theta x-\ln A(\theta)\}\le +\infty.
\ee
Note that we change notations from $a$ to $\theta$ for compatibility
with expressions later on. 
The function $I(x)$ is convex non-decreasing, with $I(0)=0$ and
$\lim_{x\rightarrow +\infty}I(x)=+\infty$ (we assume that $A(a)$ is finite for some
$a>0$).

One obtains
\be
\prob\left(\frac 1n\sum_{k=1}^nX_k\ge x\right)\le e^{-nI(x)}.
\label{intro:standardub}
\ee
When $I(x)$ is strictly positive
then  an outcome larger than $x$ is a large deviation and its probability
decays exponentially fast in $n$.  

\section{Deformed logarithmic and exponential functions}

Fix $q$ satisfying $0<q<2$, $q\not=1$.
The $q$-deformed logarithm is defined by \cite {TC94,NJ11}
\be
\ln_q(u)=\frac{1}{1-q}\left(u^{1-q}-1\right)
\quad u>0.
\ee
In the limit $q=1$ it reduces to the natural logarithm $\ln u$.
The inverse function is the $q$-deformed exponential.
It is defined on the whole of the real axis by
\be
\exp_q(u)=\left[1+(1-q)u\right]_+^{1/(1-q)}\le +\infty.
\ee
Here, $[u]_+$ denotes the positive part of $u$.
Note that $\exp_q(\ln_q(u))=u$ holds for all $u>0$.
However, $\ln_q(\exp_q(u))$ may differ from $u$ when $\exp_q(u)$ diverges
or vanishes.

For further use we mention that
\be
\exp_q(u)\exp_{2-q}(-u)
&=&\left(\frac{\left[1+(1-q)u\right]_+}{\left[1+(1-q)u\right]_+}\right)^{1/(1-q)}\cr
&=& 1,
\label{def:qqstar}
\ee
whenever $1+(1-q)u>0$.

The following two properties are used later on.

\begin{proposition}
\label{prop:logconv}
 The function $\exp_q(x)$ is log-concave when $q<1$ and log-convex when $q>1$.
\end{proposition}

\beginproof
Let $f(x)=\ln\exp_q(x)$. Its first derivative equals
\be
f'(x)&=&\left[\exp_q(x)\right]^{q-1}\cr
&=&\frac{1}{\left[1+(1-q)x\right]_+}.
\ee
This function is decreasing when $q<1$ and increasing when $q>1$.
\endproof

\begin{proposition}
\label{prop:prod}
 Let $0<q<1$ and let $q^*=2-q$.
 Then one has for all $a>0$ and $b>0$ that
 \be
 \exp_q(a+b)\le \exp_q(a)\exp_q(b)
 \label{def:q}
 \ee
and
\be
 \exp_{q^*}(-a-b)\ge \exp_{q^*}(-a)\exp_{q^*}(-b).
  \label{def:qstar}
 \ee
 
\end{proposition}

The proof is straightforward.
Note that equalities hold in the case $q=q^*=1$.

The $q$-deformed exponential distribution is defined on the positive
axis and has $\exp_q(-ax)$ as its tail distribution.
Hence the probability density is
\be
f_q(x)&=&a\left(\exp_q(-ax)\right)^q,
\quad x\ge 0,\cr
&=&a\left[1-(1-q)ax\right]_+^{q/(1-q)}.
\ee
When $0<q<1$ then the distribution has a compact support,
namely
\be
\left[0, \frac 1{a(1-q)}\right].
\ee
On the other hand, when $1<q<2$ then it has a fat tail
\be
f_q(x)\sim\frac 1{[(q-1)ax]^{q/(q-1)}}.
\ee
These two cases are rather different.
Therefore we will treat them separately.
However, in order to avoid confusion we restrict
in what follows the values of the parameter $q$ to the interval $[0,1]$
and use $q^*$ to denote values in the range between 1 and 2.
In fact, this convention has been followed already in the previous proposition.

\section{The case of a compact support}

\subsection{A deformed inequality}

The Markov inequality implies the following analogue of (\ref {intro:ub}).

\begin{proposition}
\label {prop:lb}
 Let be given i.i.d.~random variables $X_1,X_2,\cdots,X_n$.
One has for all $x$ and for all $a>0$ for which $(1-q)ax<1$
\be
\prob\left(\frac 1n\sum_{k=1}^nX_k \ge x\right)
\le \left[\exp_{q}(-ax)\right]^n A^n(a),
\label{lb:mainineq}
\ee
with
\be
A(a)=\Eo\exp_{q^*}(aX_1).
\ee
\end{proposition}

\beginproof
Because $\exp_{q^*}$ is log-convex one has
\be
\ln\exp_{q^*}\left(\frac 1n\sum_{k=1}^naX_k\right)\le
\frac 1n\sum_{k=1}^n\ln\exp_{q^*}\left(aX_k\right).
\ee
This can be written as
\be
\Io_{\sum_{k=1}^nX_k\ge nx}\,\left[\exp_{q^*}\left(ax\right)\right]^n
&\le& \left[
\exp_{q^*}\left(\frac 1n\sum_{k=1}^naX_k\right)\right]^n\cr
&\le&
\prod_{k=1}^n\exp_{q^*}\left(aX_k\right).
\ee
Here, $\Io_c$ denotes the indicator function which equals 1 when $c$ is satisfied
and vanishes otherwise.
Take the expectation. This gives
\be
\prob\left(\frac 1n\sum_{k=1}^nX_k \ge x\right)
\left[\exp_{q^*}\left(ax\right)\right]^n
\le A^n(a).
\ee
The latter can be written as (\ref {lb:mainineq}).
\endproof

We will see in an example later on that as a bound 
the above result is less sharp than  (\ref {intro:ub}).

\subsection{Legendre structure}

Introduce now a parameter $\theta$ defined by
\be
\theta=\left[\Eo\exp_{q^*}(aX_1)\right]^{1-q}a.
\ee

\begin{lemma}
\label{lem:leg}
$\theta$ is a strictly increasing function of $a$
on the open interval of $a$-values for which  
$0<\Eo\exp_{q^*}(aX_1)<+\infty$.
\end{lemma}

\beginproof
One calculates
\be
\frac{\upd\theta}{\upd a}
&=&\left[\Eo\exp_{q^*}(aX_1)\right]^{1-q}\cr
& &+(1-q) \left[\Eo\exp_{q^*}(aX_1)\right]^{-q}
\Eo\left(\left[\exp_{q^*}(aX_1)\right]^{q^*}X_1\right) a\cr
&=&\left[\Eo\exp_{q^*}(aX_1)\right]^{-q}
\Eo\left[\exp_{q^*}(aX_1)\right]^{q^*}\cr
&>& 0.
\ee

\endproof

A consequence of this lemma is that the functional dependence $\theta(a)$
may be inverted to $a(\theta)$. Hence we can define a function $\Phi(\theta)$
by
\be
\Phi(\theta)=\ln_{q}\Eo\exp_{q^*}(aX_1).
\label{lb:Phidef}
\ee
Note that $a\downarrow 0$ implies $\theta=0$.
Let
\be
{\overline\theta}&=&\sup_{a>0}\theta(a)\le+\infty.
\label{lb:overtheta}
\ee
Then $\Phi(\theta)$ is defined for $0<\theta<\overline\theta$.

\subsection{A Theorem}

The Proposition \ref{prop:lb} can now be reformulated as follows.

\begin{theorem}
Let be given  i.i.d.~random variables $X_1,X_2,\cdots,X_n$.
Fix $q$ such that $0<q<1$ and let $q^*=2-q$.
Assume that $\Eo\exp_{q^*}(aX_1)$ is finite for small positive $a$.
Then one has for all $x$ that
 \be
\prob\left(\frac 1n\sum_{k=1}^nX_k \ge x\right)
\le \left[\exp_{q}(-I(x))\right]^n,
\label{lb:ldt}
\ee
with the rate function $I(x)$ given by
\be
 I(x)=\sup_{0<\theta<\overline\theta}\{\theta x-\Phi(\theta)\}.
 \ee
 The function $\Phi(\theta)$ is defined by (\ref {lb:Phidef}).
 The range $(0,\overline\theta)$ is defined by (\ref {lb:overtheta}).
\end{theorem}

\beginproof
A short calculation shows that the r.h.s.~of (\ref {lb:mainineq})
can be written as
\be
\left[\exp_{q}(\Phi(\theta)-\theta x)\right]^n.
\ee
In this expression $\theta$ has an arbitrary value in $(0,\overline\theta)$.
The proof then follows by taking the infimum over $\theta$.

\endproof

\subsection{Example: the uniform distribution}

Consider for instance a random variable $X$ uniformly distributed on the interval $[0,1]$.
A short calculation gives
\be
A(a)&\equiv&\Eo\exp_{q^*}(aX)\cr
&=&\frac{1}{qa}\left[\left\{\exp_{q^*} (a)\right\}^{q}-1\right].
\ee
This yields
\be
\theta=aA^{1-q}=a\left[\frac{\{\exp_{q^*}(a)\}^q-1}{qa}\right]^{1-q}
\ee
and
\be
\Phi(\theta)&=&\ln_{q}A\cr
&=&\frac{1}{1-q}\left[\frac{\theta}{a}-1\right].
\ee
A short calculation shows that the quantity $\Phi-\theta x$ is minimal when
$a=0$ or $a$ is a solution of
\be
 \exp_{q^*}(a)=1+\frac{a}{1-a+axq}.
 \label{ex1:opt}
\ee
A series expansion for small values of $a$ yields
\be
\Phi(a)-\theta x=\left(\frac 12-x\right)a+{\rm O}(a^2).
\ee
This shows that $I(x)\not=0$ whenever $x>1/2$.
Hence, in this case (\ref {ex1:opt}) has a useful solution.
Note that $a<1/(1-q)$ is needed to keep $\exp_{q^*}(a)$ finite.

\begin{figure}[!t]
	\centering
	\includegraphics[width=.9\textwidth]{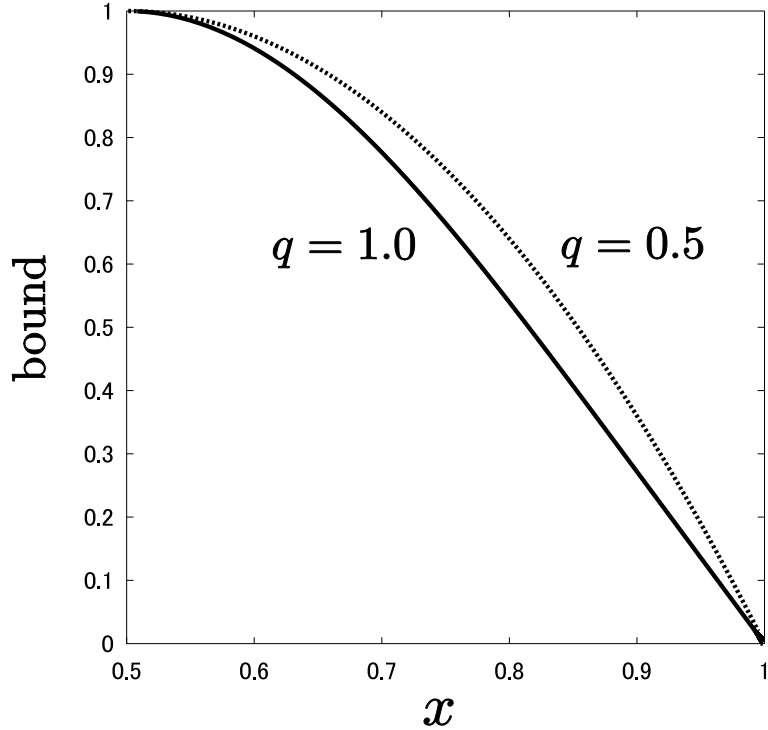}
	\caption{Upper bounds for the probability that $X_1$ is larger than $x$ 
	given the uniform distribution on the interval $[0,1]$.
	From top to bottom
	the curves correspond with $q=1/2$
	and $q=1$ (standard case).}
	\label {ex1:bounds}
\end{figure}

Take for instance $q=1/2$. This gives $A=2/(2-a)$,
$\theta=a\sqrt{A}$
and
$\Phi=2(\theta/a-1)$.
The minimum is obtained for $a=0$ or $a=4(x-1/2)/x$.
The latter requires $1/2<x<1$.
One obtains
\be
I(x)
&=&2-4\sqrt{x(1-x)},
\quad \frac 12<x<1
\ee
The final result is then
\be
\prob\left(\frac 1n\sum_{k=1}^nX_k \ge x\right)
\le \left[4x(1-x)\right]^{n},
\quad \frac 12\le x\le 1.
\label{ex1:ub}
\ee
Note that this result can be written as
\be
\prob\left(\frac 1n\sum_{k=1}^nX_k \ge x\right)
\le e^{-nI_1(x)},
\quad \frac 12\le x\le 1
\ee
with
\be
I_1(x)=\ln 2x +\ln 2(1-x).
\ee
One can show numerically that the bound (\ref {ex1:ub}) is less sharp than the one
obtained by the standard inequality ($q=1$). However, (\ref {ex1:ub}) has the advantage
of being expressed in a closed form.
See the Figure \ref {ex1:bounds}.

\section{The $q^*$-deformed exponential distribution}
\subsection{Definition}
Fix $q$ between 0 and 1, as before, and let $q^*=2-q$.
Let
\be
\eta(x)&=&\exp_{q^*}(-x)\quad x\ge 0,\cr
&=&1\quad x\le 0.
\ee
Let $X$ be a random variable distributed according to the distribution $f(x)$ given by
\be
f(x)&\equiv&\frac{\upd\,}{\upd x}(1-\eta(x))\cr
&=&\left[\exp_{q^*}(-x)\right]^{q^*}\quad \mbox{if } x> 0,\cr
&=&0\qquad\qquad\mbox{if } x<0.
\label{ref:distr}
\ee
Then one has
\be
\prob\left(X \ge x\right)=\eta(x).
\ee
This distribution is a special case of the Lomax distribution \cite {LKS54} and hence of a
type-II Pareto distribution. Its first moment exists and is given by
\be
\Eo X=\frac 1q.
\ee

An important property of this distribution is the following.
Note that in the case of the exponential distribution (this is the $q=1$-limit)
it holds with equality.

\begin{proposition}
\label{prop:prodineq}
 \be
 \eta(a)\eta(b)\le \eta (a+b),
 \quad \mbox{for all } a>0, b>0.
 \ee
\end{proposition}

\beginproof
One can write
\be
\eta(a)\eta(b)
&=&
\frac{1}{[1+(1-q)a]^{\frac{1}{1-q}}}\,
\frac{1}{[1+(1-q)b]^{\frac{1}{1-q}}}\cr
&=&
\frac{1}{[1+(1-q)(a+b)+(1-q)^2ab]^{\frac{1}{1-q}}}\cr
&\le&
\frac{1}{[1+(1-q)(a+b)]^{\frac{1}{1-q}}}\cr
&=&\eta(a+b).
\ee
\endproof

\subsection{Sums of i.i.d.~variables}

The law of large numbers holds for the distribution (\ref {ref:distr}).
Hence one can expect that some form of a large deviation principle should hold.

Consider a sequence of i.i.d.~variables $X_1$, $X_2$, $\cdots$, $X_n$,
all distributed according to $f(x)$ given by (\ref {ref:distr}) and introduce
tail distributions $\eta_n$ defined by
\be
\eta_n(x)=\prob\left(\frac 1n\sum_{k=1}^nX_k \ge x\right).
\ee
These functions will be used later on in the formulation of a large deviation estimate.
They satisfy the inequalities
\be
\left[\eta(x)\right]^n\le \eta_n(x)\le 1.
\ee
The lower bound can be improved easily. Indeed, one has

\begin{proposition}
\label{prop:ineq}
For all $x>0$ is
\be
& &
1-[1-\eta(nx)]^n
\le \eta_n(x).
\label{ref:ineqs}
\ee
\end{proposition}

This result is a special case of Proposition \ref {prop:sums:lb} found below.

It turns out to be very difficult to obtain a sharp upper bound, valid
for arbitrary values of $n$. Therefore we go immediately over to an asymptotic analysis.

\subsection{Asymptotic analysis}

For large values of $x$ the functions $\eta_n(x)$ satisfy the relation
$\eta_n(x)\sim n\eta(nx)$. This property is known to be
equivalent with sub-exponentiality \cite{TJL75}.
From
\be
\eta(x)\sim \left[\frac{1}{(1-q)x}\right]^{1/(1-q)}
\ee
then follows that
\be
n^{q/(1-q)}\eta_n(x)\sim\eta(x)\quad\mbox{ as }x\rightarrow\infty.
\ee
This suggests that for large $n$ and for $x>\Eo X_1=1/q$
the expression\\
 $n^{q/(1-q)}\eta_n(x)$ remains bounded when $n$ tends to infinity.
This turns out to be correct, as discussed below.

From the lower bound (\ref{ref:ineqs}) follows immediately that
\be
\liminf_{n\rightarrow\infty}n^{q/(1-q)}\eta_n(x)
\ge\left[\frac{1}{(1-q)x}\right]^{\frac{1}{1-q}},
\quad x>0.
\ee
Indeed, one has
\be
\nonumber
n^{q/(1-q)}\left(1-[1-\eta(nx)]^n\right)
&\sim&n^{1+q/(1-q)}\eta(nx)\cr
&\sim&n^{1+q/(1-q)}\left[\frac{1}{(1-q)nx}\right]^{1/(1-q)}\cr
&=&\left[\frac{1}{(1-q)x}\right]^{1/(1-q)}.
\ee
In particular, this result implies that the standard Large Deviation Principle is not satisfied.
For the asymptotic upper bound we have to appeal on the mathematical analysis
originally started by Heyde \cite {HCC67,HCC67b} and Nagaev \cite {NAV69,NAV69a}.
The $q$-exponential distribution
belongs to the class of distributions they consider.
As a consequence, one has the following result.

\begin{proposition}
\label{prop:expdistrasym}
 For all $x>\Eo X_1$ and for $n$ tending to $\infty$ is
 \be
 \eta_n(x)\sim n\eta(n(x-\Eo X_1))
 \sim \frac{1}{n^{\frac{q}{1-q}}}\left[\frac{1}{x-\Eo X_1}\right]^{\frac{1}{1-q}}.
 \label{qdef:asym}
 \ee 
\end{proposition}

\beginproof
See for instance Theorem A in \cite {STJ01}.
\endproof

\section{The case of a fat tail}

\subsection{The deformed inequality}

The result of Proposition \ref {prop:lb} is not valid for $q>1$ because the proof uses that
$\exp_{q^*}$ is log-convex. However, a slightly different result is obtained
using Proposition \ref {prop:prod} instead of \ref {prop:logconv}.

\begin{proposition}
\label {prop:lo}
 Let be given positive i.i.d.~random variables $X_1,X_2,\cdots,X_n$.
One has for all $x>0$ and for all $a>0$
\be
\prob\left(\frac 1n\sum_{k=1}^nX_k \ge x\right)
\le \exp_{q^*}(-anx) A^n(a),
\label{lo:mainineq}
\ee
with
\be
A(a)=\Eo\exp_{q}(aX_1).
\ee
\end{proposition}

\beginproof
Because $a>0$ and $\exp_q$ is an increasing function one has
\be
\Io_{\{\sum_{k=1}^nX_k\ge nx\}}\exp_{q}\left(anx\right)
&\le& \exp_{q}\left(a\sum_{k=1}^nX_k\right).
\ee
Now use Proposition \ref {prop:prod} to obtain
\be
\Io_{\{\sum_{k=1}^nX_k\ge nx\}}\exp_{q}\left(anx\right)
&\le&\prod_{k=1}^n\exp_{q}\left(aX_k\right).
\ee
Take the expectation. This gives, with the help of the i.i.d.~property
of the random variables,
\be
\prob\left(\sum_{k=1}^nX_k\ge nx\right)
\exp_{q}\left(anx\right)
&\le& \left[\Eo\exp_{q}\left(aX_1\right)\right]^n.
\ee

This result can be written as (\ref {lo:mainineq}).

\endproof

We will use this result only for $n=1$. The factor $A^n(a)$
in the r.h.s.~of (\ref {lo:mainineq}) diverges exponentially fast
and prohibits sharp estimates in the limit of large $n$.

\subsection{Sums of i.i.d. variables} 

The lower bound (\ref {ref:ineqs})
is a special case of the following easy lower bound.

\begin{proposition}
\label{prop:sums:lb}
Let be given  i.i.d.~random variables $X,X_1,X_2,\cdots,X_n$,
all following the same probability distribution $f(x)$.
Let ${\overline F}(x)$ denote the corresponding tail distribution.
Then one has
\be
1-[1-{\overline F}(nx)]^n\le \prob\left(\frac 1n\sum_{k=1}^nX_k \ge x\right).
\ee 
\end{proposition}

\beginproof
One has
\be\nonumber
\prob\left(\frac 1n\sum_{k=1}^nX_k \ge x\right)
&=&\int\upd x_1\,f(x_1)\cdots\int\upd x_n\,f(x_n)\Io_{\{\sum x_k\ge nx\}}\cr
&=&n\int\upd x_1\,f(x_1)\int^{x_1}\upd x_2\,f(x_2)\cr
& &
\cdots\int^{x_1}\upd x_n\,f(x_n)\Io_{\{\sum x_k\ge nx\}}.
\ee
To see this note that one may assume that one of the variables, say $x_1$,
is larger than the others.
Next use that it is sufficient that $x_1$ is larger than $nx$
to obtain
\be
\prob\left(\frac 1n\sum_{k=1}^nX_k \ge x\right)
&\ge&n\int_{nx}\upd x_1\,f(x_1)\int^{x_1}\upd x_2\,f(x_2)
\cdots\int^{x_1}\upd x_n\,f(x_n)\cr
&=&n\int_{nx}\upd x_1\,f(x_1)\left[1-{\overline F}(x_1)\right]^{n-1}\cr
&=&1-[1-{\overline F}(nx)]^n.
\ee
\endproof

The Proposition \ref {prop:lo} is used to obtain an upper bound.

\begin{proposition}
\label{theorem:fat}
Let be given  i.i.d.~random variables $X,X_1,X_2,\cdots,X_n$.
Fix $q$ such that $0<q<1$ and let $q^*=2-q$.
Let $A(a)=\Eo\exp_{q}(aX)$. 
Assume $A(a)$ is finite for all $a>0$.
If $x>\frac{1}{a}\ln_q(A(a))$ then
\be
 \prob\left(\frac 1n\sum_{k=1}^nX_k \ge x\right)
\le \eta_n(y).
\label{prop:ub2}
\ee
with
\be
y=aA^{1-q^*}x-\ln_{q^*}A.
\label {thm2:y}
\ee
\end{proposition}

\beginproof
Note that the condition
\be
x>\frac{1}{a}\ln_q(A(a))
\ee
implies that $y$ defined by (\ref {thm2:y}) is positive.
It also implies that
$x>\Eo X_1$. To see this use the concavity of the function $\ln_q$.

Consider the probability distribution
\be
g(y)&=&aA(a)\left[\exp_{q^*}(-ay)\right]^{q^*}, \quad y >y_0,\cr
&=&0\quad\mbox{otherwise,}\ee
with $y_0$ given by $ay_0=\ln_q(A(a))$.
Let $Y$ be a random variable with pdf $g(y)$.
Then one has
\be
\prob\left(Y\ge x\right)&=&A(a)\exp_{q^*}(-ax).
\ee
The Proposition \ref {prop:lo} then shows that 
\be
\prob\left(X \ge x\right)
\le \prob\left(Y \ge x\right).
\ee
This implies that
\be
\prob\left(\frac 1n\sum_{k=1}^nX_k \ge x\right)
\le
\prob\left(\frac 1n\sum_{k=1}^nY_k \ge x\right),
\ee
where the $Y_k$ are i.i.d.~with pdf $g(y)$.
Now write
\be
\prob\left(\frac 1n\sum_{k=1}^nY_k \ge x\right)
&=&\int_{y_0}^\infty\upd y_1\,g(y_1)\cdots\int_{y_0}^\infty\upd y_n\,g(y_n)\,
\Io_{\{\sum_{k=1}^ny_k \ge nx\}}.\cr
& &
\ee
Introduce new integration variables
\be
x_k=-\ln_{q^*}A\exp_{q^*}(-ay_k).
\ee
This gives
\be
\prob\left(\frac 1n\sum_{k=1}^nY_k \ge x\right)
&=&\int_{y=y_0}^\infty\upd x_1\,\left[A(a)\exp_{q^*}(-ay_1)\right]^{q^*}
\cdots\cr
& &\int_{y=y_0}^\infty\upd x_n\,\left[A(a)\exp_{q^*}(-ay_n)\right]^{q^*}\,
\Io_{\{\sum_{k=1}^ny_k \ge nx\}}\cr
&=&\int_{0}^\infty\upd x_1\,\left[\exp_{q^*}(-x_1)\right]^{q^*}
\cdots
\int_{0}^\infty\upd x_n\cr
& &\times
\left[\exp_{q^*}(-x_n)\right]^{q^*}\,
\Io_{\{\sum_{k=1}^n\ln_{q^*}\left(\eta(x_k)/A\right) \le -nax\}}.\cr
& &
\ee

Hence the inequality reduces to 
\be
\prob\left(\frac 1n\sum_{k=1}^nX_k \ge x\right)
&\le& 
\left(\prod_{k=1}^n\int^{0}_{+\infty}\upd \eta(x_k)\right)
\Io_{\{\sum_{k=1}^n\ln_{q^*}(\eta(x_k)/A)\le -nax\}}.\cr
& &
\label{thm2:temp}
\ee
Note that
\be
& &\sum_{k=1}^n\ln_{q^*}(\eta(x_k)/A)\le -nax\cr
&\leftrightarrow&
\sum_{k=1}^n\frac{1}{1-q^*}\left[(\eta(x_k)/A)^{1-q^*}-1\right]\le -nax\cr
&\leftrightarrow&
\sum_{k=1}^n(\eta(x_k)/A)^{1-q^*}\ge n[1+(1-q)ax]\cr
&\leftrightarrow&
\sum_{k=1}^n(\eta(x_k))^{1-q^*}\ge nA^{1-q^*}[1+(1-q)ax]\cr
&\leftrightarrow&
\sum_{k=1}^n\left[1+(1-q)x_k\right]_+\ge nA^{1-q^*}[1+(1-q)ax].
\ee
The $x_k$ are positive integration variables. Therefore the condition becomes
\be
& &n+(1-q)\sum_{k=1}^nx_k\ge nA^{1-q^*}[1+(1-q)ax]\cr
&\leftrightarrow&
\sum_{k=1}^nx_k\ge \frac{n}{1-q}\left[A^{1-q^*}[1+(1-q)ax]-1\right]\cr
&\leftrightarrow&
\sum_{k=1}^nx_k\ge ny
\ee
with $y$ given by (\ref{thm2:y}).
(\ref {thm2:temp}) can now be written as (\ref{prop:ub2}).

\endproof

\subsection{A Large Deviation Result} 

The above result can now be combined with the known asymptotics of the function
$\eta_n(x)$ as found in Proposition \ref {prop:expdistrasym}.
This yields
\be
\prob\left(\frac 1n\sum_{k=1}^nX_k \ge x\right)
&\le& \eta_n(y)
\ee
with $y=aA^{1-q^*}x-\ln_{q^*}A$ and
\be
\eta_n(y)
&\sim& n\eta\left(n\left(y-\frac{1}{q}\right)\right).
\ee

Introduce now a parameter $\theta$ defined by
\be
\theta=\frac{a}{A^{1-q}(a)}.
\ee
It takes values in the range $(0,\overline\theta)$ with
\be
\overline\theta=\lim_{a\rightarrow\infty}\frac{a}{A^{1-q}(a)}\le\infty.
\ee

\begin{lemma}
 $\theta$ is an increasing function of $a$.
\end{lemma}

\beginproof
Note that
\be
\frac{\upd\,}{\upd a}
\exp_q(aX)&=&\left[\exp_q(aX)\right]^q\cr
&=&\frac{1}{(1-q)a}\left\{\exp_q(aX)-\left[\exp_q(aX)\right]^q\right\}
\ee
so that
\be
\frac{\upd A}{\upd a}
&=&\frac{\upd\,}{\upd a}\Eo \exp_q(aX)\cr
&=&\frac{1}{(1-q)a}\left\{A(a)-\Eo \left[\exp_q(aX)\right]^q\right\}.
\ee
This is used in the following calculation
\be
\frac{\upd\theta}{\upd a}
&=&\frac{\theta}{a}\left[1-(1-q)\frac{a}{A(a)}\frac{\upd\,}{\upd a}\right]\cr
&=&
\frac{\theta}{aA(a)}\Eo\left[\exp_q(aX_1)\right]^q,
\ee
which is a positive quantity.

\endproof

This allows us to define a function $\Phi(\theta)$ by
\be
\Phi(\theta)&=&\ln_{q^*}(A(a)).
\ee
We use it to write
\be
n\eta\left(n\left(y-\frac{1}{q}\right)\right)
&\sim&
\frac{1}{n^{\frac{q}{1-q}}}
\frac{1}{\left[(1-q)\left(aA^{1-q^*}(a)x-\ln_{q^*}(A)-\frac 1q\right)\right]^{\frac{1}{1-q}}}\cr
&=&\frac{1}{n^{\frac{q}{1-q}}}
\frac{1}{\left[(1-q)\left(\theta x-\Phi(\theta)-\frac 1q\right)\right]^{\frac{1}{1-q}}}\cr
&\sim&n
\exp_{q^*}(n[\frac 1q+\Phi(\theta)-\theta x]).
\ee
The parameter $\theta$ can still be chosen freely.
Hence we can optimize the asymptotic bound by taking the infimum over $\theta>0$.
The results obtained so far can be summarized in the following theorem.

\begin{theorem}
\label{theorem:fattails}
 Let be given  i.i.d.~random variables $X_1,X_2,\cdots,X_n$.
Fix $q$ such that $0<q<1$ and let $q^*=2-q$.
Let $A(a)=\Eo\exp_{q}(aX)$. 
Assume $A(a)$ is finite for all $a>0$.
Introduce a parameter $\theta$,
a constant $\overline\theta$ and a function $\Phi(\theta)$ in the way described above.
Introduce a rate function $I(x)$ by
\be
I(x)=\sup_{\theta}\{\theta x-\Phi(\theta):\,0<\theta<\overline\theta\}.
\ee
There exist functions $\xi_n(x)$ such that
\be
\prob\left(\frac 1n\sum_{k=1}^nX_k \ge x\right)
&\le& \xi_n(x)
\ee
with the property that
\be
\xi_n(x)
\sim
n\exp_{q^*}(\frac nq-nI(x)).
\ee
\end{theorem}

\subsection{Example}

The Student's t-distribution is given by
\be
f(x)=\frac 1{\sqrt{\nu\pi}}\frac{\Gamma(\frac {\nu+1}2)}{\Gamma(\frac {\nu}2)}
\left(1+\frac{x^2}{\nu}\right)^{-(\nu+1)/2}.
\ee
Its variance diverges when $\nu\le 2$.The $q$-moment generating function
$A(a)=\Eo \exp_{q}(aX)$ converges when $q<1-1/\nu$.

\begin{figure}[!t]
	\centering
	\includegraphics[width=.9\textwidth]{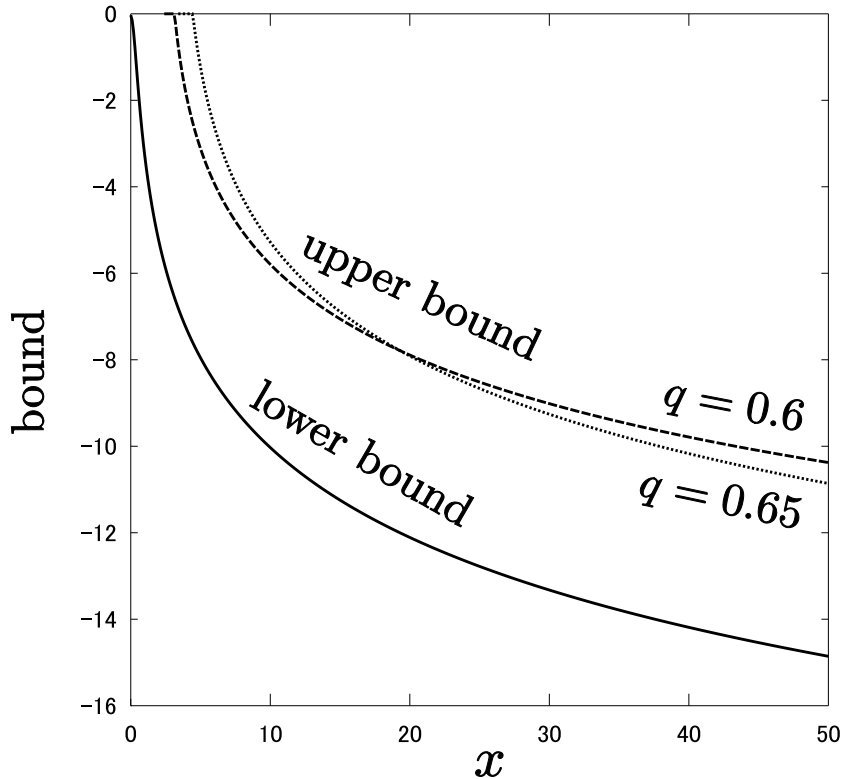}
	\caption{Lower bound (full line) and asymptotic upper bounds as a function of $x$
	for the tail distribution of the sum
	of 5 i.i.d.~variables distributed according to student t with $\nu=3$.
	The vertical axis shows the logarithm of the bounds.
	The parameters of the upper bounds are $q=0.6$ and $q=0.65$, respectively.
        In both cases is $a=5$.
	}
	\label {ex2:bounds}
\end{figure}

Take for instance $\nu=3$.
The probability distribution is
\be
f(x)=\frac{2}{\pi\sqrt 3}\,\frac{1}{(1+\frac 13x^2)^2}.
\ee
The tail distribution is
\be
\overline F(x)
&\equiv&\int_x^\infty\upd y \,f(y)\cr
&=&\frac{2}{\pi\sqrt 3}\int_x^\infty\upd y\,
\frac{1}{(1+\frac 13y^2)^2}\cr
&=&\frac 12-\frac{1}{\pi}\arctan\frac{x}{\sqrt 3}-\frac{\sqrt 3}{\pi}\frac{x}{3+x^2}.
\ee
The lower bound behaves for large $n$ as
\be
1-\left[1-\overline F(nx)\right]^n
\sim
n\overline F(nx)\sim \frac{2\sqrt 3}{\pi n^2x^3}.
\ee
Comparison of the latter with (\ref{qdef:asym}) suggests to take $q=2/3$
when evaluating the upper bound.
This is indeed the limiting value for the existence of the deformed generating function
$A(a)$. We therefore plot in Fig. \ref {ex2:bounds} upper bounds for 
different values of $q$ slightly less than $q=2/3$.
In addition, instead of numerically minimizing over $\theta$ to obtain the rate function
$I(x)$, upper bounds for a fixed value of $\theta$, or equivalently of $a$, are plotted.
These are given by
\be
n\exp_{q^*}\left(\frac nq+\frac{n}{1-q}-\frac{n}{1-q}\frac{1+(1-q)ax}{A(a)^{1-q}}\right).
\ee

\section{Summary and Discussion} 

Our starting point is an application of the Markov inequality to
variables of the form $\exp_q(aX)$, where $\exp_q$ is the $q$-deformed exponential function
and $a>0$ is a free parameter.
We use this to obtain an upper bound for sums of i.i.d.~variables.
In the case of a probability distribution with a compact support this leads to an elegant
formalism which however is less powerful than the standard treatment.
In the case of probability distributions with a fat tail we proceed by comparison with the
$q^*$-deformed exponential distribution with $q^*=2-q$ and $0<q<1$. 
Large deviation estimates for the latter distribution are obtained from 
results found in the literature. Our main result is Theorem \ref {theorem:fattails}.
It uses the analogy between the $q$-deformed and the standard exponential function
to formulate a large deviation principle for distributions with a fat tail.

Is it worthwhile to introduce $q$-deformed exponential functions in the
theory of large deviations? We know that there is no fundamental reason for their usage.
The L\'evy distributions are the appropriate tools for studying distributions
with a fat tail. However, they are rather complicated. The main advantage of
the $q^*$-deformed exponential distribution is therefore its simplicity.
The possibility of proceeding by analogy with the conventional approach
is a plus point.
We interpret the standard theory of large deviations as
a comparison of arbitrary distributions with the exponential distribution.
Theorem \ref {theorem:fattails} is based on a comparison of fat-tailed distributions
with the $q^*$-deformed exponential distribution.

The present work is a first attempt to use $q$-deformed exponential functions in the
context of large deviation theory. The main theorem is probably not optimal.
The two examples serve as an illustration and fall short of showing the full potential
of the present approach. Further work is therefore needed.

\section*{Acknowledgement}

This work was done during the second author's sabbatical stay in Antwerp
University, which was financially supported by JSPS KAKENHI Grant Number
25540106.

\end{document}